\documentclass[aoas,MSNbibl,nameyear,dvips]{arximspdf}
\usepackage{graphicx}
\doi{10.1214/14-AOAS714} 
\volume{8}
\issue{2}
\pubyear{2014}
\firstpage{956}
\lastpage{973}
\setattribute{copyright}{owner}{In the Public Domain}


\begin{document}
\begin{frontmatter}

\title{Adjusting models of ordered
multinomial outcomes for nonignorable nonresponse
in the occupational employment statistics survey}
\runtitle{Adjusting models of ordered multinomial
outcomes}

\begin{aug}
\author[A]{\fnms{Nicholas J.} \snm{Horton}\corref{}\thanksref{aff1,t1}\ead[label=e1]{nhorton@amherst.edu}},
\author[B]{\fnms{Daniell} \snm{Toth}\thanksref{aff2}\ead[label=e2]{toth.daniell@bls.gov}}
\and
\author[B]{\fnms{Polly} \snm{Phipps}\thanksref{aff2}\ead[label=e3]{phipps.polly@bls.gov}}
\runauthor{N.~J. Horton, D. Toth and P. Phipps}

\thankstext{t1}{Funded by the American Statistical
Association, National Science Foundation and Bureau of Labor Statistics
(ASA/NSF/BLS)
Fellowship program during the period 2011--2013.}
\affiliation{Amherst College\thanksmark{aff1} and Bureau of Labor
Statistics\thanksmark{aff2}}
\address[A]{N. J. Horton\\
Department of Mathematics and Statistics\\
Amherst College\\
Amherst, Massachusetts 01002\\
USA\\
\printead{e1}}
\address[B]{D. Toth\\
P. Phipps\\
Office of Survey Methods Research\\
Bureau of Labor Statistics, Suite 1950\\
Washington, DC 20212\\
USA\\
\printead{e2}\\
\phantom{E-mail:\ }\printead*{e3}}
\end{aug}

\received{\smonth{5} \syear{2013}}
\revised{\smonth{1} \syear{2014}}

%
\begin{abstract}
An establishment's average wage, computed from administrative wage data,
has been found to be related to occupational wages.
These occupational wages are a primary outcome variable for the Bureau
of Labor Statistics
Occupational Employment Statistics survey.
Motivated by the fact that nonresponse in this survey is associated
with average wage
even after accounting for other establishment characteristics,
we propose a method that uses the administrative data for
imputing missing occupational wage values due to nonresponse.
This imputation is complicated by the structure of the data.
Since occupational wage data
is collected in the form of counts of employees in predefined wage
ranges for each occupation,
weighting approaches to deal with nonresponse
do not adequately adjust the estimates for certain domains of estimation.
To preserve the current data structure, we propose a method to impute
each missing establishment's wage interval count data as an ordered
multinomial random variable
using a separate survival model for each occupation.
Each model incorporates known auxiliary information for each
establishment associated
with the distribution of the occupational wage data, including
geographic and industry characteristics.
This flexible model allows the baseline hazard to vary by occupation
while allowing predictors
to adjust the probabilities of an employee's salary falling within the
specified ranges.
An empirical study and simulation results suggest that the method
imputes missing OES wages
that are associated with the average wage of the establishment in a way
that more closely resembles the observed association.\looseness=1
\end{abstract}

%

\end{frontmatter}
\newpage
\section{Introduction}\label{sec1}

Every large survey has to contend with nonresponding units.
Estimates are commonly adjusted by rescaling each unit's sample weight
proportionally to the inverse of its response probability.
Response probabilities are modeled conditionally on available auxiliary
information
[see, e.g., \citet{KimKim07}].
For example, the linear estimator $\sum_{i \in S} w_i\mathbf{Y}_i $
for a population parameter based on the dependent variable $\mathbf
{Y}_i $
using data from the sample $S$ would be adjusted as
\[
\sum_{i \in S_R} p(\mathbf{x}_i)^{-1}
w_i\mathbf{Y}_i,
\]
where $\{w_i\}_{i=1}^N$ is a set of predefined fixed sample weights,
$S_R$ is the subset of responding units in the sample,
and $p(\mathbf{x}_i)$ is the estimated probability of unit $i$ responding
given the auxiliary information $X_i=x_i$.

There are a number of methods for adjusting the weights.
These include using weighting classes [\citet{Lit82}],
post-stratification [\citet{HolSmi79}],
calibration [\citet{Kot06}] and nearest neighbor approaches [\citet{CheSha00}].
All of the weight adjustment methods aim to account for
missing data of the nonrespondents by scaling up the data of responding units.
Therefore, these weighting adjustments implicitly impute missing data
as a linear combination of the observed data.

Weight adjustments can reduce bias introduced due to the missing information
if the auxiliary variables used to calculate nonresponse probabilities
are predictive of nonresponse.
The adjustments can even reduce the mean squared error of an estimator
if these variables
are also associated with the outcome variable [\citet{LitVar05}].
Adjusted estimators will be unbiased if the nonresponse mechanism
is missing at random (MAR) given the auxiliary information and
the response propensity model is correctly specified [\citet{Rub76};
\citet{LitRub02}].

However, when the probability of response is associated with outcome
variables even after conditioning
on all auxiliary variables, the estimator could still be biased even
after adjusting the weights for nonresponse.
For example, \citet{Schetal11}
report on a study of bone mineral density using scan data,
where most of the missing data was a direct result of a subject's
characteristics such as body mass
index (BMI) and age.
In fact, subjects over a certain age or with a BMI over a specific
limit were excluded from the scan (by medical necessity).
Therefore, all observed scan data came from younger subjects and/or
subjects with lower BMI
compared to some subjects with unobserved data.
Since these characteristics are also highly associated with the outcome
variable (bone mineral density),
adjustment for this type of missingness requires a unverifiable model
(explicitly or implicitly)
to extrapolate from the observed data [\citet{ChaKot08}; \citet{KotCha10}].

In this article, we propose to adjust the occupational wage data from the
Bureau of Labor Statistics (BLS) Occupational Employment Statistics
survey (OES)
using auxiliary variables obtained from administrative data to reduce
bias due to unit nonresponse.
Among the available variables is the average wage per employee for each
establishment.
\citet{PhiTot12} demonstrated that this average wage tended to be lower
for responding establishments than for nonrespondents, even after conditioning
on the other auxiliary variables.
In addition, the work of \citet{Gro91} and \citet{LanSalSpl}
suggests an establishment's average wage is highly associated with
occupational wages at that establishment.

Adjusting the occupational wage data for nonresponse is complicated by
the structure of the data.
The OES collects total counts of employees at each establishment in twelve
predefined ordered wage ranges
for each occupation.
By collecting the wage interval data in this manner
the OES data yield quantile estimates of wages for each occupation as
well as averages.
Thus, this data structure has more utility than aggregated totals of
employment and wages,
but adjusting for missing values becomes more difficult.

Indeed, if only mean wages were collected, the data could be
adjusted using one of the weighting methods enumerated previously.
However, due to the structure of the data and because
there are certain domains for which occupational wages estimates are
produced that contain
nonresponding units with a higher establishment-level average wage per
employee than all responding units,
we argue that any weighting approach to adjust for nonresponse will
lead to biased domain estimates.

To preserve the current data structure, we propose a method to impute
each missing establishment's wage interval count data as an ordered
multinomial random variable,
using a separate survival model for each occupation.
Each model incorporates known auxiliary information associated
with the distribution of the occupational wage data.
Section~\ref{oes} introduces the OES survey data, which provides the motivation
for the proposed model-based method of adjustment.
Section~\ref{impute} describes the new method and explains how it
incorporates the administrative data.
Section~\ref{results} compares the imputed data produced using the new procedure
with the existing method used by the OES and includes results of a
simulation study.
The code and data sets for these models are available at the
Bureau of Labor Statistics through the external research program.

\section{The OES occupational wage data}\label{oes}
The BLS Occupational Employment Statistics (OES) survey, an
establishment survey, measures occupational employment and wage rates for
around 800 occupations in the Standard Occupational Classification
(SOC) system
by industry for states and territories in the United States.
Estimates are produced at the national, state and metropolitan
statistical area levels
using data from approximately 1.2 million sampled establishments.

The sample is drawn from the Quarterly Census of Employment and Wages (QCEW),
a frame of about 9 million establishments across all nonfarm industries.
The frame contains administrative data on a number of variables for
every establishment,
including total employment for each of the three months in the quarter;
total wages paid during that quarter ($\mathsf{WAGE}$);
data-defined groups of industries ($\mathsf{IND}$) defined by the six digit
North American Industry Classification System (NAICS) code;
the metropolitan or nonmetropolitan area where the establishment is
located ($\mathsf{MSA}$),\vadjust{\goodbreak}
an indicator of whether the MSA is in
the largest of six size categories ($\mathsf{MSACATT6}$),
and whether or not the establishment is part of a national
multi-establishment firm ($\mathsf{MULTI}$).
The average quarterly wage per employee,
%
\begin{equation}
\mathsf{AVEWAGE}=\mathsf{WAGE}/\mathsf{EMPL},
\end{equation}
is calculated for every establishment (including nonrespondents) using
the frame data.
The variable $\mathsf{EMPL}$ is the average reported employment over
the three months in the quarter.
The establishments are sampled using a probability proportional to size
(p.p.s.) design
that is stratified by industry and area.


A responding establishment reports wages for the OES survey by
occupational code.
The data is reported by the establishment $i$ by listing the number of
employees 
with a given occupational code 
that have hourly wages contained in a given interval $I_l=[a_l,   b_l]$,
for each of twelve wage intervals $l=1, \ldots, 12$.
Table~\ref{table:oesdata} illustrates the tabular form of the
occupational data collected by the OES
for an establishment.

%
\begin{table}
\tabcolsep=0pt
\caption{Example of the data collected for OES
occupational wage survey from establishment $i$.
The number of employees $e_{icl}$, with SOC code $c=1,   \ldots,
C_i$, that have an hourly wage contained
in interval $I_l=[a_l,   b_l]$, for $l=1,   \ldots,  12$, is
collected for each of the $C_i$ occupations
represented at establishment $i$.
The twelve wage intervals are the same for every occupation code, but
vary across states
(these differences are limited to the first two wage intervals and are
driven by the state's minimum wage).
For example, the November 2006 panel, the lower bound of hourly-wage
interval one, $a_1$, ranges by state
from \$6.35 to \$8.42 per hour, while the lower bound of the second
interval, $a_2$,
ranges from \$8.42 to \$10.61 per hour.
The last wage interval $I_{12}$ is open, $[a_{12}, \infty)$.
These wage intervals are regularly adjusted for inflation, so they
change over time}\label{table:oesdata}
\begin{tabular*}{\textwidth}{@{\extracolsep{\fill}}lcccccccccccc@{}}
\hline
\textbf{SOC} &$\bolds{I_{1}}$ &$\bolds{I_{2}}$ &$\bolds{I_{3}}$ &$\bolds{I_{4}}$
&$\bolds{I_{5}}$ &$\bolds{I_{6}}$
&$\bolds{I_{7}}$&$\bolds{I_{8}}$&$\bolds{I_{9}}$& $\bolds{I_{10}}$
& $\bolds{I_{11}}$ & $\bolds{I_{12}}$\\
\hline
1 & $e_{i11}$ & $e_{i12}$ & $e_{i13}$ & $e_{i14}$ & $e_{i15}$ &
$e_{i16}$ & $e_{i17}$ &
$e_{i18}$ & $e_{i19}$ & $e_{i110}$ & $e_{i111}$ & $e_{i112}$\\
2 & $e_{i21}$ & $e_{i22}$ & $e_{i23}$ & $e_{i24}$ & $e_{i25}$ &
$e_{i26}$ & $e_{i27}$ &
$e_{i28}$ & $e_{i29}$ & $e_{i210}$ & $e_{i211}$ & $e_{i212}$\\
\vdots&&&&&& \vdots&&&
&&&\vdots\\
$c$ & $e_{ic1}$ & $e_{ic2}$ & $e_{ic3}$ & $e_{ic4}$ & $e_{ic5}$ &
$e_{ic6}$ & $e_{ic7}$ &
$e_{ic8}$ & $e_{ic9}$ & $e_{ic10}$ & $e_{ic11}$ & $e_{ic12}$\\
\vdots&&&&&& \vdots&&&&&&\vdots\\
$C_{i}$ & $e_{iC_{i}1}$ & $e_{iC_{i}2}$ & $e_{iC_{i}3}$ &
$e_{iC_{i}4}$ & $e_{iC_{i}5}$ &
$e_{iA_{i}6}$ & $e_{iA_{i}7}$ &
$e_{iC_{i}8}$ & $e_{iC_{i}9}$ & $e_{iC_{i}10}$ & $e_{iC_{i}11}$ &
$e_{iC_{i}12}$\\
\hline
\end{tabular*}
\end{table}

This wage interval data is used to
produces estimates of the total employment, mean wage, mean salary, as
well as estimates for the hourly and annual 10th, 25th, 50th, 75th and
90th percentile wages
for every occupation.
Details on how these estimates are computed using interval data are presented
in Piccone and Hesley (\citeyear{PicHes}).
OES also publishes a mean relative standard error for the total
employment and mean wage estimates.
In order to retain the same level of utility of the OES data,
any imputation procedure must produce counts for each of twelve wage intervals.\vadjust{\goodbreak}


\section{Imputing ordered multinomial wage data}\label{impute}

Though the OES achieves a high overall response rate (78\%, one of the
highest of any BLS establishment survey),
bias in the estimates due to nonresponse could still be a problem.
This is particularly true in smaller subdomains if the estimates are
not properly adjusted.
Currently the OES uses a nearest-neighbor type imputation procedure to
account for
missing occupational wage data due to nonresponse.
For each nonresponding establishment, the list of occupations to impute,
as well as the proportion of employees in each occupation,
is taken from a donor establishment selected from a group of identified
neighbors
based on industry and location.
The missing wage interval data is then imputed by taking an average of
all nearby responding units' data,
where neighboring establishments are identified based on geography,
industry, size and ownership status.
More details regarding the imputation and estimation procedures can be
found in
the OES State Operations Manual [\citet{bure11}].

The current method adjusts for unit nonresponse by replacing the
missing unit's data with
an average of the data from responding establishments that are similar
to the nonresponding unit.
However, when all the responding units in a neighborhood have smaller
values of $\mathsf{AVEWAGE}$
than a nonresponding unit, weighting-up wage data of the responding
units will not
adequately account for the missing wage data for occupations that are associated
with the $\mathsf{AVEWAGE}$ of an establishment.
For example, Table~\ref{ex1} gives a hypothetical example (similar to
situations confronted by the OES)
of five establishments' wage data for a given occupation in the form
collected by OES.
Three of the establishments responded, while two establishments did not respond.

%
\begin{table}
\caption{Hypothetical example of occupational wage
data for five establishments in a given
domain defined by industry and geographic area.
Each row gives the establishment's id, total number of employees with
the given occupational code,
counts of the number of employees in each of the twelve wage intervals,
the average wage per employee $\mathsf{AVEWAGE}$, and the response indicator,
where $R=1$ means that the establishment responded to the OES survey.
The wage interval counts are not observed (italicized) for the two
establishments with $R=0$.
In this example, only five establishments employ workers with this
occupational code in this industry and area,
and the two establishments with the highest $\mathsf{AVEWAGE}$
(establishments 2 and 4) did not respond to the survey}
\label{ex1}
\begin{tabular*}{\textwidth}{@{\extracolsep{\fill}}lccccccccccccccc@{}}
\hline
\textbf{id} & \textbf{tot} & \textbf{1} & \textbf{2} & \textbf{3} & \textbf{4} &
\textbf{5} & \textbf{6} & \textbf{7} & \textbf{8} & \textbf{9} & \textbf{10} &
\textbf{11} & \textbf{12} & \textbf{AVEWAGE}
&\multicolumn{1}{c@{}}{$\bolds{R}$} \\
\hline
1 & \phantom{0}8 & 0 & 0 & 0 & 0 & 0 & 1 & 3 & 1 & 2 & 1 & 0 & 0 & 15,981 & 1 \\
2 & \emph{10} & \emph{0} & \emph{0} & \emph{0} & \emph{0} & \emph{0} &
\emph{0} & \emph{0} & \emph{1} & \emph{2} & \emph{2} & \emph{1} & \emph
{4} & 23,364 & 0 \\
3 & \phantom{0}4 & 0 & 0 & 0 & 0 & 1 & 2 & 0 & 0 & 0 & 1 & 0 & 0 & \phantom{0,}8420 & 1 \\
4 & \phantom{0}\emph{7} & \emph{0} & \emph{0} & \emph{0} & \emph{0} & \emph{0} &
\emph{0} & \emph{0} & \emph{0} & \emph{1} & \emph{1} & \emph{3} & \emph
{2} & 27,343 & 0 \\
5 & \phantom{0}8 & 0 & 0 & 0 & 0 & 0 & 0 & 2 & 0 & 4 & 2 & 0 & 0 & 15,058 & 1 \\
\hline
\end{tabular*}
\end{table}

In this example, no weight-based method will work using these establishments
because replacing either of the missing establishments' data with any
linear combination of the responding
establishments' data will lead to inflated cell counts at the lower
level of wage categories and have
cell counts of zero at the highest wage categories.
This is not representative of the nonresponding establishments' data.
Any method for adjusting for nonresponse should be able to adequately
adjust the shape of the induced
empirical cumulative distribution function (ECDF).

\begin{figure}

\includegraphics{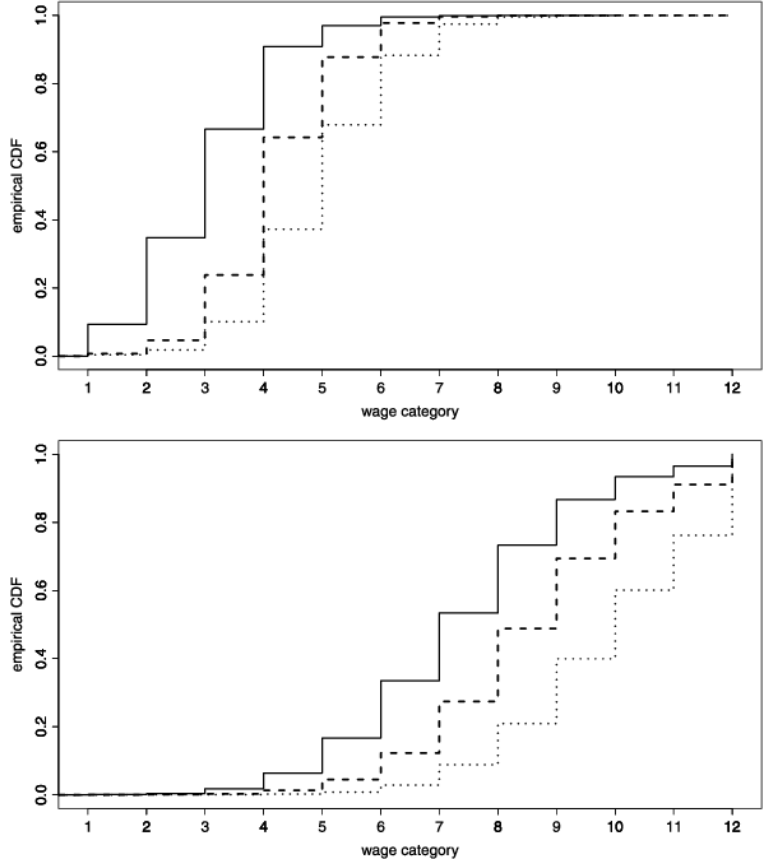}

\caption{Empirical cumulative distribution function
(CDF) for customer service representatives
(top) and general managers (bottom) stratified by tertiles of $\mathsf
{AVEWAGE}$.
The wage of an employees at an establishment in the highest tertile of
$\mathsf{AVEWAGE}$
(dotted line) is more likely to fall in a higher wage interval than
that of an employee
at an establishment in a lower tertile (middle${} = {}$dashed, lowest${} = {}$solid).
This is shown to be true for both the lower wage earning occupation,
customer service representatives,
as well as the higher wage earning occupation, general managers.}
\label{fig:ecdf}
\end{figure}

To illustrate, consider the distribution of wage categories from
respondents for two common
occupations: customer service representatives and general managers.
Figure~\ref{fig:ecdf} displays the ECDF
for the 12 wage categories using all observed data from establishments reporting
on customer service representatives (top) and from establishments
reporting on general managers (bottom)
stratified by tertiles of the average wage distribution ($\mathsf
{AVEWAGE}$) computed from the frame data.
An employee's wage at an establishment in the highest tertile of
$\mathsf{AVEWAGE}$ is more likely to fall in a
higher wage interval than that of an employee at an establishment in a
lower tertile.
Note that the overall distribution is shifted to the right in
comparison with the customer service representatives,
with higher average wages for general managers reflecting the generally
higher pay for this occupation.

We propose a method that extrapolates from $\mathsf{AVEWAGE}$ of each
establishment,
as well as incorporates other important establishment characteristics
using a model of the ECDF.
The proposed model of the probability that an individual employee
(in that establishment for a given occupation)
falls in each of the 12 wage categories is modeled as a flexible
``time-to-event'' (survival) model.
We then sample from the resulting distribution to impute missing OES
occupational wage values.

We use a Cox proportional hazards regression for discrete failure times
to model this wage distribution [\citet{Cox72}].
For a given establishment with characteristic variables $\mathbf{X}$,
let $I$ be the interval in which the employee's salary falls for $I= \{
1, \ldots, 12\}$.
The model for the hazard is given by
%
\begin{equation}
\lambda(l) = \frac{P(L=l)}{P(L \geq l)} =\lambda_0(l)\exp{\bolds{
\beta}'\mathbf{X}}, \label{eq:fullmod}
\end{equation}
where $\lambda_0(l)$ is the baseline hazard function,
\begin{eqnarray*}
\beta' &=& (\beta_0, \beta_1, \ldots,
\beta_9),
\\
\mathbf{X}'&=&{ \bigl(1, \mathbf{X}_1,
\mathbf{X}_1^2, \mathbf{X}_2,
\mathbf{X}_2^2, \mathbf{X}_3,
\mathbf{X}_1\mathbf{X}_3, \mathbf{X}_4,
\mathbf{X}_5, \mathbf{X}_6 \bigr)},
\end{eqnarray*}
where
$\mathbf{X}_1=\mathsf{AVEWAGE}$,
$\mathbf{X}_2=\log(\mathsf{EMPL})$,
$\mathbf{X}_3=\mathsf{SINGLE}$,
$\mathbf{X}_4=\mathsf{MSACATT6}$,
$\mathbf{X}_5=\mathsf{BMSA}$, and
$\mathbf{X}_6=\mathsf{MULTI}$.

Since there are no censored outcomes, all employees must end up in one
of the 12 wage categories.
The \citet{Efr77} estimator is used to account for ties, since there are only
twelve possible intervals.
\citet{The} states that this approximation is more accurate than the
Breslow when dealing with tied death times as well as computationally efficient.
Use of the exact conditional likelihood is not computationally
tractable for this example.

Quadratic terms are included for $\mathsf{AVEWAGE}$ and $\log(\mathsf
{EMPL})$ to
account for the quadratic relationship between these variables and
occupational wages.
We also adjust the establishments employing only a single employee
within a given occupation differently,
because the relationship between the occupational wage and
$\mathsf{AVEWAGE}$ was observed to be different than establishments
with multiple employees within that occupation.
Likewise, the occupational wages of employees of establishments located
in large MSAs had
a different relationship to $\mathsf{AVEWAGE}$ than those in smaller MSAs.
Therefore, we added two occupation specific variables $\mathsf{SINGLE}$
and $\mathsf{BMSA}$.
The variable $\mathsf{SINGLE}=1$, if the establishment has only a
single employee in that occupation
and $\mathsf{BMSA}$ is an identifier of any MSAs where there are many
($\geq$250)
establishments employing people in the given occupation (see \hyperref[app]{Appendix}
for details on handling MSAs).

The baseline hazard function $\lambda_0(t)$ is a data-driven occupation
specific step function,
which can jump at each of the wage categories $t$.
This is estimated using the empirical distribution of wage category counts.
The model is stratified on classes defined using $\mathsf{IND}$
so that a different baseline hazard function $\lambda_0(t)$ is
estimated for each industry class
(see \hyperref[app]{Appendix} for details on how the industry classes are defined).
We fit the model using the survival package in R [\citet{The}],
accounting for clustering within each establishment.

Given the establishment-level variables for a nonresponding establishment,
the survival model parameter estimates can be used to generate
predicted probabilities
that an employee with the given occupation falls into each of the 12
possible wage categories.
We impute the missing data for the establishment by taking a random
draw from a multinomial distribution
with those probabilities for each employee at the establishment within
that occupation.

This approach is attractive because the model for the hazard
of falling into a lower wage category can extrapolate on average wage
while including the other establishment-level characteristics in the
imputation process.
Due to the model's flexibility, it allows the occupation specific
baseline hazard
(of falling into a lower wage category) to be unspecified, with
predictors controlling the shape within the
constraints of the model.

Currently, the OES uses a jackknife procedure to estimate sampling
error, with additional variance components due to the categorization of
wage data. In order to account for variability due to imputation,
imputations could be resampled within each jackknife replication pool.

%

\section{Evaluating the proposed imputation method}\label{results}

One difficulty with assessing nonresponse adjustment methods
and their impact on a real survey is that the true missing values are unknown.
In this section we attempt to evaluate the proposed imputation
procedure by
comparing the imputed values to observed values,
considering the impact of the adjustment on estimates like those
produced by the OES,
as well as testing the procedure using a simulation with known response
probabilities.

\subsection{Comparison with existing OES imputation procedure}

\begin{figure}

\includegraphics{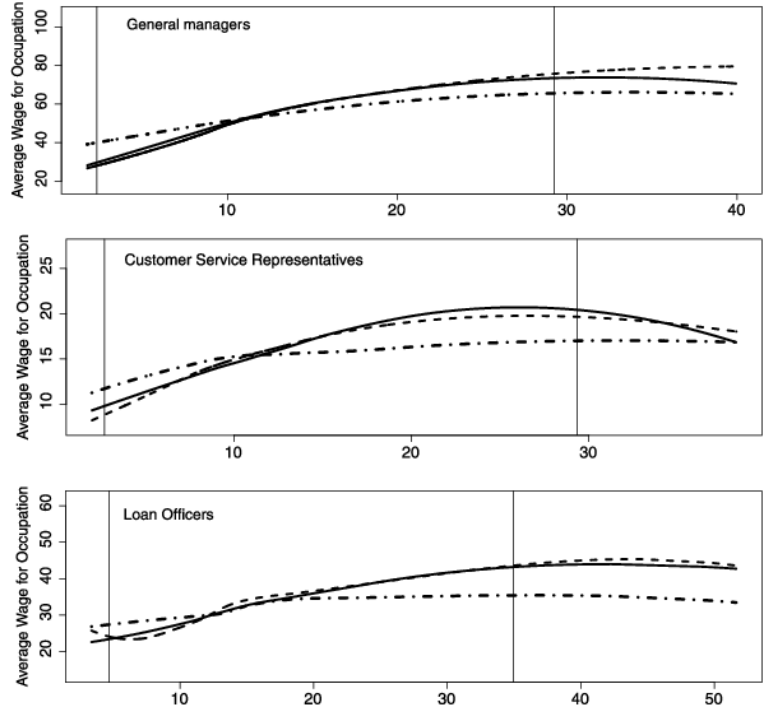}

\caption{Average occupational wage for each establishment computed using imputed
values from the new and current imputation methods compared to the
observed data.
The y-axis gives the average occupational wage for each establishment
computed using
imputed data from the new method (solid line), imputed data from the
current OES method (dash-dot line) and the observed wage data (dashed line).
The three average occupational wages are plotted over the average
overall wage for each establishment, $\mathsf{AVEWAGE}$, computed from
the administrative record data.
The computed occupational average wages are displayed for three occupations:
general managers (top), customer service representatives (middle) and
loan officers (bottom).
The comparisons show that the occupational average wages computed using
the imputed
data from the new method match the observed average wages more closely
than the
average wages using the imputed data from the old method.
The two vertical lines indicate the location of the 2.5th and the
97.5th percentiles of $\mathsf{AVEWAGE}$.}
\label{fig:custsvcimpute}
\end{figure}

We start by comparing the relationship
of the imputed values and the predictor variables to that of the
observed values.
Obviously, if the missing values are due to nonignorable nonresponse,
there is no reason the
imputed values should have the same relationship to predictor variables
as the observed values,
but it would seem desirable all the same [\citet{AbaGelLev08}].
Figure~\ref{fig:custsvcimpute} displays the predicted relationship between
the occupational mean wage and $\mathsf{AVEWAGE}$
for general managers, customer service representatives and loan officers.
The dashed line indicates the observed relationship between $\mathsf
{AVEWAGE}$ and wage within an occupation
for the respondents, while the solid line represents imputations using
the proposed method and
the dash-dot line represents imputations using the existing OES method
over $\mathsf{AVEWAGE}$.
All three lines are smoothed estimates of the functional relationship
among occupational wage and
the variable $\mathsf{AVEWAGE}$.
The wage curves demonstrate an increasing relationship with respect to
$\mathsf{AVEWAGE}$.
Also, the proposed new method (solid line) more closely parallels the
relationship between $\mathsf{AVEWAGE}$
and the observed wage for that occupation when compared to the existing
OES imputation method (dash-dot line)
for all three occupations shown here.
This was the case for all occupations tested in an empirical study.

Next we consider the impact of the new imputation procedure on OES
estimates by comparing
estimates from data adjusted with the proposed imputation procedure to
those using the existing OES procedure.
Table~\ref{table:compare} gives estimates for the mean and 75th percentile
for six occupations using the observed and imputed values.
National estimates are compared as well as those for two sub-domains,
one based on industry (Commercial Banking) and
another by area (Chicago MSA).
Six representative occupations
were chosen because they
are common within the two sub-domains, represent a wide range of high
to low paying occupations,
and have varying proportions of nonresponse.

\begin{table}
\tabcolsep=0pt
\def\arraystretch{0.95}
\caption{Comparison of OES estimates (hourly mean
wage in dollars and 75th percentile
for hourly wage) for the
occupational categories: janitor, customer service representative, loan officer,
computer information systems manager, general manager and lawyer.
The estimates are computed using the completed data set using the
current and then from the new imputation methods.
Estimates are computed for these occupations overall as well as for the
commercial banking industry
and for the Chicago area MSA}
\label{table:compare}
\begin{tabular*}{\textwidth}{@{\extracolsep{\fill}}lccccccc@{}}
\hline
& \textbf{Occupation} & \textbf{Janitors} & \textbf{Cust. svc}. &
\textbf{Loan officer} & \textbf{CIS mgr.} & \textbf{Gen.
mgr.} & \textbf{Lawyer} \\
\hline
Overall & $n$ & 25,936 & 21,089 & 2,629 & 7,788 & 57,292 & 4,081 \\
& prop. obs. & 73.2\% & 67.5\% & 74.1\% & 59.0\% & 70.3\% & 68.9\% \\
& OES mean & \$10.80 & \$14.70 & \$31.70 & \$54.10 & \$51.00 & \$61.40
\\
& new mean & \$11.00 & \$14.90 & \$31.90 & \$54.20 & \$50.90 & \$62.60
\\
& OES 75th & \$12.60 & \$17.00 & \$37.20 & \$62.20 & \$64.20 & \$78.30
\\
& new 75th & \$12.50 & \$17.00 & \$37.60 & \$62.20 & \$62.20 & \$78.20
\\[3pt]
Commercial & $n$ & 177 & 788 & 990 & 178 & 603 & 51 \\
banks & prop. obs. & 76.8\% & 81.2\% & 81.5\% & 75.3\% & 75.5\% &
80.4\% \\
& OES mean & \$9.70 & \$14.90 & \$31.50 & \$63.90 & \$52.70 & \$72.70
\\
& new mean & \$10.40 & \$15.20 & \$31.90 & \$64.90 & \$54.40 & \$72.40
\\
& OES 75th & \$10.50 & \$16.60 & \$37.80 & \$75.20 & \$66.00 & \$85.80
\\
& new 75th & \$10.60 & \$17.00 & \$39.40 & \$76.00 & \$69.30 & \$86.80
\\[3pt]
Chicago & $n$ & 486 & 535 & 49 & 296 & 1059 & 104 \\
MSA area & prop. obs. & 58.4\% & 60.6\% & 55.1\% & 49.7\% & 57.3\% &
43.3\% \\
& OES mean & \$13.80 & \$17.40 & \$33.60 & \$56.30 & \$55.50 & \$49.70
\\
& new mean & \$13.00 & \$19.10 & \$35.10 & \$56.00 & \$55.20 & \$52.10
\\
& OES 75th & \$15.30 & \$20.70 & \$37.60 & \$65.60 & \$70.50 & \$61.30
\\
& new 75th & \$13.10 & \$21.80 & \$42.10 & \$67.20 & \$73.30 & \$68.50
\\
\hline
\end{tabular*}    \vspace*{-3pt}
\end{table}

From Table~\ref{table:compare}, we see that even at the national level,
mean estimates of occupational wages
in the empirical investigation changed by several percent using the new
imputation method
for certain occupations (e.g., janitors and lawyers).
However, as one would expect, larger differences occur in both the mean
and quantile estimates
in most occupations in the less aggregated sub-domains.
This is a function of both the relatively high nonresponse rates for
some sub-domains compared to the
national response rate for these occupations
and in the distribution of $\mathsf{AVEWAGE}$ values in these
sub-domains.\looseness=1

\subsection{Assessing behavior of the imputations when the true
missingness mechanism is known}

We undertook a simulation study to evaluate the performance of the
method under a variety of models for
nonresponse of the occupational wages in realistic settings using the
complete cases as our population.
Another difficulty in assessing the impact of the new method on OES
published estimates is
that we do not have access to the exact algorithm currently used to
impute values.
Therefore, we cannot directly compare the imputed values from the new
method to new values imputed by the OES.
Instead, we attempt to assess the importance of including $\mathsf
{AVEWAGE}$ in the imputation model.
For each of three scenarios, two models were fit: the full model
described by model (\ref{eq:fullmod}),
which we call \texttt{FULL}, as well as a simplified model that did not
control for
$\mathsf{AVEWAGE}$, which we call \texttt{NO-AVEWAGE}.

Let $R$ denote an indicator that the occupational wage data are observed.
Missingness was set to approximate the 20\% rate of being unobserved,
with the following logistic model:
\begin{eqnarray*}
\mathrm{ logit} P(R=0) &=& \alpha_0 + \alpha_1 \log(
\mathsf{EMPL}) + \alpha_2 \mathsf{MSACATT6} + \alpha_3
\mathsf{AVEWAGE} \\
&&{}+ \alpha_4 \mathsf{OCCWAGE}.
\end{eqnarray*}
Missingness of the occupational wage data was imposed using one of
three mechanisms:
%
\begin{enumerate}[\texttt{NINR}]
\item[\texttt{MAR1}] Missingness was Missing at Random (MAR)
in the sense of \citet{LitRub02}.
More specifically, it depends on $\log(\mathsf{EMPL})$ and being in the
largest MSA size category, $\mathsf{MSACATT6}$
($\alpha_0 = -2.89$; $\alpha_1= 0.105$; $\alpha_2= 2.42$; $\alpha
_3=\alpha_4=0$).

\item[\texttt{MAR2}] Same as \texttt{MAR1} plus missingness also depends on
$\mathsf{AVEWAGE}$
($\alpha_0 = -3.39$; $\alpha_1= 0.105$; $\alpha_2= 2.42$; $\alpha
_3=0.0000262$; $\alpha_4=0$).

\item[\texttt{NINR}] Missingness depends only on the unobserved
occupational wage\break
$\mathsf{OCCWAGE}$ ($\alpha_0 = -2.39$; $\alpha_1=\alpha_2= \alpha
_3=0$; $\alpha_4=0.02$).
\end{enumerate}

In the simulation we use computer information systems managers as our
example occupation.
Using the data from the 4595 responding establishments that employed
people in this occupation,
we generate 250 partially observed data sets for each of the three
scenarios and two models.
For each of the generated data sets, the average as well as 75th
percentile of income was calculated
for the missing data using the imputed values using \texttt{FULL} and \texttt{NO-AVEWAGE}.
Each estimate was compared to the average and 75th percentile of the
true values.

\begin{figure}[!t]

\includegraphics{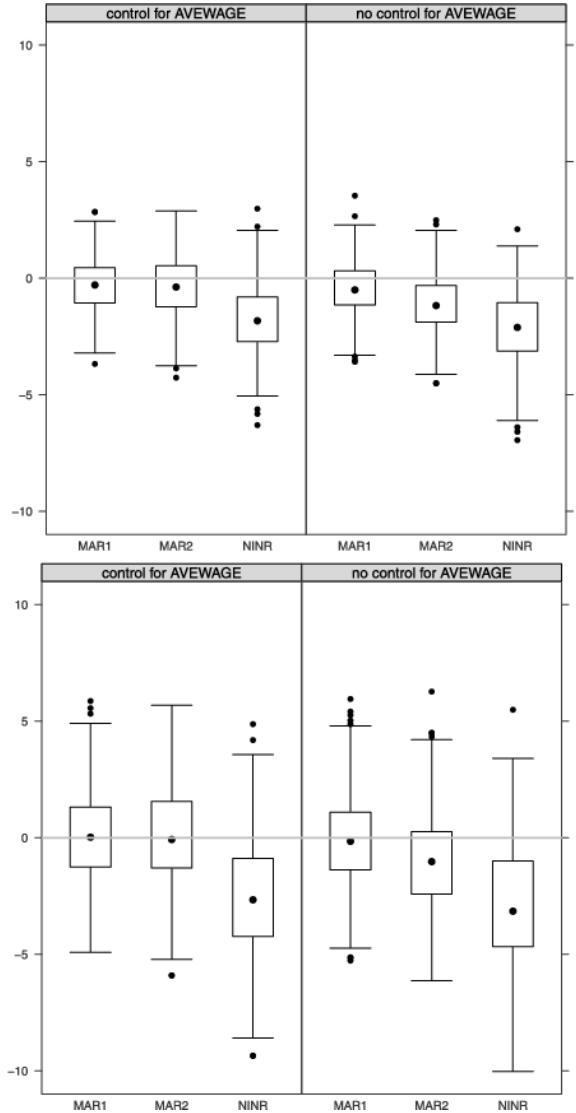}

\caption{Boxplot of the 250 differences between
estimates using imputed values
and those using the true values from the simulation study
(using computer information system managers as the occupation).
The estimates are for the nonresponding units only.
The top figure displays the differences between mean occupational wage estimates
and the bottom figure shows the differences between the 75th percentile
estimates.
Results are shown for the three different missing-data mechanisms used:
\texttt{MAR1}, \texttt{MAR2} and \texttt{NINR}.
\texttt{MAR1} depends only on employment size and being in large MSA,
\texttt{MAR2} is the same as \texttt{MAR1} plus association with $\mathsf{AVEWAGE}$,
while \texttt{NINR} depends only on the unobserved mean occupational wage.}
\label{fig:simres}
\end{figure}

Figure~\ref{fig:simres} displays the differences between estimates
using the imputed values
and the true values for missing establishments for each of the three scenarios.
Neither model was biased for the \texttt{MAR1} scenario.
Including $\mathsf{AVEWAGE}$ in the model minimized bias in the \texttt{MAR2} scenario,
as would be expected given that this was a key predictor of missingness.
Both models yielded bias in the \texttt{NINR} scenario, however, there was
modest improvement
using the model which controlled for $\mathsf{AVEWAGE}$.
These results highlight the importance of broadening the set of
variables included in the
imputation model as a way to make the missing at random assumption more
tenable [\citet{ColSchKam01}].

\section{Discussion}\label{discussion}

We proposed a flexible, yet explicit model for imputing missing
occupational wages into categories.
The model incorporates administrative wage data at the establishment
level as well as geography,
industry and other establishment characteristics to account for
nonresponse bias.
Unlike a weight-based method of adjustment for nonresponse,
this imputation approach is able to extrapolate missing occupational
wage values using establishment wages.

During the evaluation of this method, it was shown that the model
generated imputations that more
closely paralleled the observed relationship
between the average wage and the observed occupational wage at that
establishment
compared to the existing OES method.
At high levels of aggregation, where the response rates for OES are
high (roughly 78\%),
the mean and quantile estimates of occupational wages were similar to
those using the existing imputation method.
However, more substantial differences were seen in estimates for
sub-domains defined by industry and MSA.

The importance of broadening the set of variables to include $\mathsf
{AVEWAGE}$ in the imputation model was also
highlighted by the results of the simulation.
Under the \texttt{NINR} simulation scenario, where nonresponse only
depended on the occupational wage,
there was some bias reduction using the model which controlled for
$\mathsf{AVEWAGE}$.
This occurred even though the simulations were based on the observed
data from responding establishments,
which tend to have lower $\mathsf{AVEWAGE}$ values than nonresponding units.

We chose a Cox proportional hazards model for our imputation process.
This has a number of attractive features, including a flexible
model for the baseline hazard (which is fit separately for each occupation
as well as important industry classes) and the ability to cluster employees
within establishment.
Other models (such as a proportional odds model or a multinomial
logistic model) might be considered
as an alternative approach, though such approaches would have to be
extended to allow stratification.
In our own evaluation of these methods, none were better than one using
a stratified Cox model.
This may be a topic for future research.

The current OES procedures utilize a hot deck imputation for employment
followed by
a weighting method to impute missing wages.
Our goal was to improve the estimation of occupational wage estimates
utilizing auxiliary information
by replacing their second-stage weighting procedure with an approach
that accounts for differences in average wage at the establishment level.
There are several options to account for errors due to the imputation.
One approach would be to re-sample imputations within each jackknife
replication pool,
while adjusting the hot deck procedure so that the total number of
employees for each unit
are sampled from an appropriate (posterior) distribution.
Further discussion and consideration of other options is a topic for
future research.

These evaluations indicate that the new method is likely to produce
imputed values that more closely
match the missing values.
This may lead to more accurate estimates of occupational wages produced
by the OES. 

\begin{appendix}\label{app}
\section*{Appendix: Details on incorporating establishment variables}

When applying a method to a large survey like the OES one usually
encounters a number of issues
concerning the data that must be handled before a method can be put
into practice.
In this section we report on issues encountered while trying to
incorporate the information from
a few of the establishment-level variables in the model and discuss how
they were handled.

\subsection{Classifying NAICS code}
Because pay rates for certain occupations depend on the industry
$(\mathsf{IND})$,
it is an important factor to include in the model.
However, there are over 1100 different NAICS codes and some
occupations only exist in a small subset of the codes.
Additionally, observed employment is very dense in some industry codes
and sparse in others,
and this pattern is occupation specific.
Therefore, including each possible $\mathsf{IND}$ as a factor in the
model is impractical.

We address this separately for each occupation code by clustering
similar NAICS codes
in order to form classes of industry codes that have relatively
homogeneous pay structures
for that occupation.
The clustering is done using a nonparametric procedure (regression tree),
recursively splitting on $\mathsf{IND}$, as a continuous variable.
This yields classes where industry codes are close to similar industries
and maintains the inherent hierarchical structure of NAICS codes.
For example, all establishments with a six-digit code starting with 52
(52XXXX) are in the super sector
``Finance and Insurance,'' while all establishments with a code in the
form 524XXX are in the
subcategory ``Insurance Carriers and Related Activities.''
The code 5241XX defines the even more refined subcategory of
``Insurance Carriers,'' while
52413X refines this to ``Reinsurance Carriers.''
Also, this method automatically splits industries where that occupation
is dense while aggregating
industries where the occupation is sparse.

In order to produce homogeneous classes, the mean of the observed wage
distribution
is used as the dependent variable in the tree regression.
The number of groups depends on the variability of the mean wages as
well as the sample size, $n$,
within each occupation.
We specified a minimum size for each industry class of 80
observed establishments employing people with that occupation code.
These classes are then used as stratification variables for the
baseline hazard function.

\begin{figure}

\includegraphics{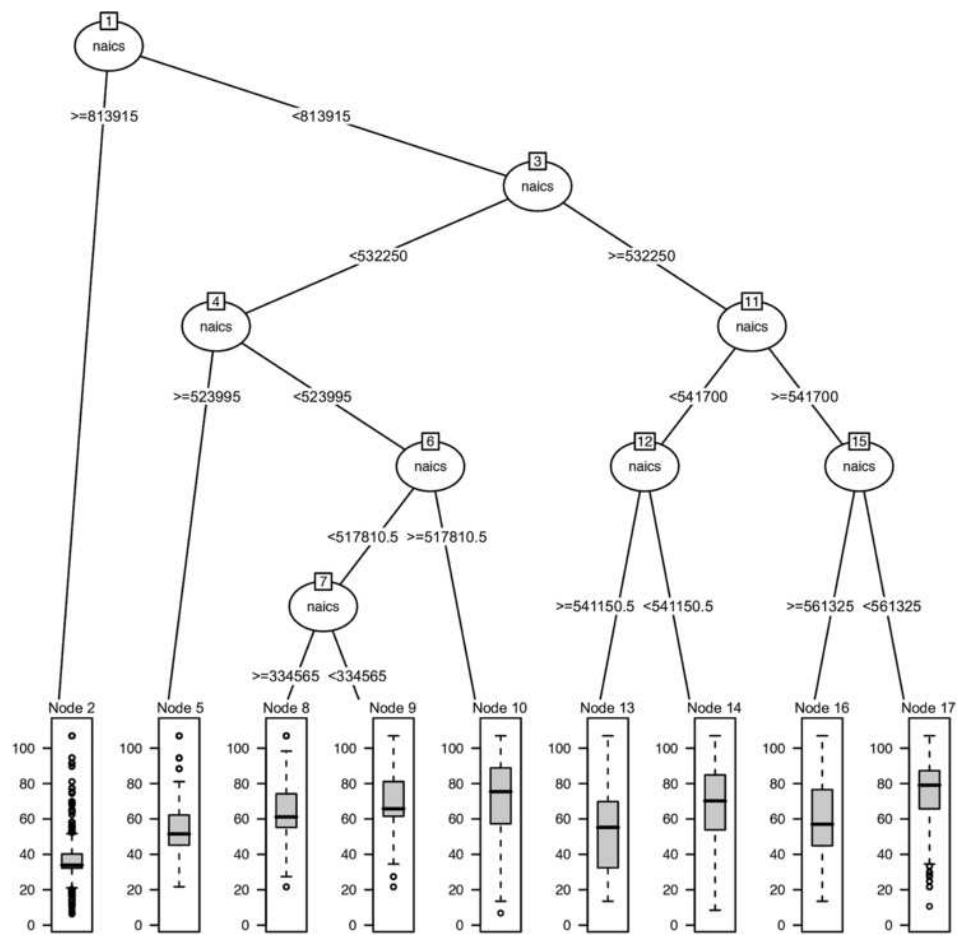}

\caption{Results from regression tree partition of
NAICS code
for lawyers using observed mean wage.
Each end node represents establishments employing lawyers in industries
within a given range of NAICS codes. The boxplot shown at the bottom of
each of the
nine end nodes displays the distribution of the average wage rate for
lawyers among these establishments.
The nodes above the end nodes show where the recursive partitioning
algorithm chose to split the industries.
Most of the splits occur on the 6-digit industry codes between 51XXXX
and 56XXXX, representing
industries which often employ lawyers and among those where wage rates
differ substantially.}
\label{fig:lawyerstree}
\end{figure}

For lawyers, with 4081 establishments in the survey (of which 2813
were observed), this
resulted in a partitioning of NAICS nodes consisting of nine distinct
nodes ranging from
a minimum size of 192 (Node~13) to the largest node (Node~14) with a
size of 1480.
The partitioning also has the intuitively appealing classification of
industries shown in
Figure~\ref{fig:lawyerstree}.

Most of the splitting occurred among the six-digit NAICS codes ranging
from 51XXXX and 56XXXX
which make up the information, financial and professional service industries.
This is expected since these are the industries in which most lawyers
are employed and for
which the salaries have the most variability.
Of the nine nodes all but three are composed primarily of industries in
this range.
One notable exception is local government (NAICS 999300), which also
contains a large number of
establishments with lawyers, dominated Node 2 (the node with the lowest
average occupational wage).\looseness=1


For comparison, there were 21,089 establishments in the survey
reporting employees
with the occupational category customer service representative, of
which 14,232 were observed.
This yielded 16 distinct NAICS classes that were also interpretable and
homogeneous with respect to average wage.

\subsection{Including MSA information}
Similarly, occupational pay rates often depend on the location of the
establishment,
therefore, metropolitan statistical area ($\mathsf{MSA}$) is an
important variable to include in the model.
But as with industry, there are too many MSAs to include each MSA as a strata.
Much of the association between occupational wage rates
and the MSA in which the establishment is located
is determined by the size of the MSA.
Therefore, for most MSAs we include an indicator in the model for
whether the establishment was in the largest of
six size categories, but for
areas with sufficiently large MSA's (at least 250 establishments
employing that occupation)
the unique MSA label was also included in the model as a factor.

\subsection{Addressing irregularities in extreme QCEW values}

The proposed method extrapolates occupational wage data at an
establishment using\break
$\mathsf{AVEWAGE}$ obtained from the administrative QCEW data.
Observed OES and QCEW data suggest a positive association between an
establishment's computed occupational
wage and $\mathsf{AVEWAGE}$ for every occupation considered during the
empirical investigation.
However, extreme values of $\mathsf{AVEWAGE}$ occur in the QCEW because
of unusually
high reported wage or low reported employment values (resulting in an
unusually large $\mathsf{AVEWAGE}$)
or low reported wage values (resulting in an unusually small $\mathsf
{AVEWAGE}$).

Extremely high reported values for wages occurring in the QCEW are
usually driven by very high bonuses
paid during the quarter or large payouts taken by the owners of the
establishment.
Unusually low employment counts (even zero), when positive wages are
reported, occur because
employment data count only workers on the payroll during the pay period that
includes the 12th day of the month, while all wages paid are reported
as wages.
Both of these lead to large $\mathsf{AVEWAGE}$ due to situations that
are unlikely to be associated
with the wage rates paid by the establishment.

Small reported earnings relative to the number of workers usually
results from a small average
number of hours worked by the employees at an establishment.
Since the QCEW does not record number of hours worked, an establishment
that has slowed down production
for a period may have a large number of employees on the payroll who
have worked minimal hours.
This situation would lead to a small reported $\mathsf{AVEWAGE}$ for
that establishment,
but would likely be unrelated to the wage rates paid by the establishment.

Despite the strong evidence of an association between an
establishment's computed occupational
wage and $\mathsf{AVEWAGE}$, this association is unlikely to hold at
the extreme tails of $\mathsf{AVEWAGE}$.
To address outliers in the values of $\mathsf{AVEWAGE}$, we recoded all values
outside the middle 98\% of the distribution.
Values between the minimum and the 1.0th percentile were recoded to be
the value of 1.0th percentile while values
between the 99.0th percentile and the maximum were recoded to the
99.0th percentile value.
This has the effect of changing relatively few of the reported values,
but still protects us from over
extrapolation.
\end{appendix}

%
\section*{Acknowledgments}
The authors would like to thank the Editor, Associate Editor and the
referees for their insightful comments made during the review process.
We gratefully acknowledge the support of Dave Byun and Michael Buso at
the BLS.
We also thank Nathaniel Schenker and John Eltinge for comments on an
earlier draft of this article and
Alana Horton for technical assistance.

Any opinions expressed in this paper are those of the author(s) and do
not constitute policy of the Bureau of Labor Statistics.


%

%


%
%





\printaddresses

\end{document}